\documentclass[final,3p,times]{elsarticle}

\usepackage{graphics}
\usepackage{graphicx}
\usepackage{epsfig}

\usepackage{amssymb}
\usepackage{amsthm}

\usepackage{lineno}
\usepackage{color}

\begin{document}

\begin{frontmatter}



\title{
Bifurcations analysis of turbulent energy cascade
}


\author{Nicola de Divitiis}

\address{"La Sapienza" University, Dipartimento di Ingegneria Meccanica e 
Aerospaziale, Via Eudossiana, 18, 00184 Rome, Italy, \\
phone: +39--0644585268, \ \ fax: +39--0644585750, \\ 
e-mail: n.dedivitiis@gmail.com, \ \  dedivitiis@dma.dma.uniroma1.it, \ \ nicola.dedivitiis@uniroma1.it}

\begin{abstract}
This note studies the mechanism of turbulent energy cascade
through an opportune bifurcations analysis of the Navier--Stokes equations, 
and furnishes explanations on the more significant characteristics of the
turbulence.
A statistical bifurcations property of the Navier--Stokes equations in 
fully developed turbulence is proposed, and a spatial representation of the
bifurcations is presented, which is based on a proper definition of the fixed
points of the velocity field.
The analysis first shows that the local deformation can be much more rapid than
the fluid state variables, then explains the mechanism of energy cascade through
the aforementioned property of the bifurcations, and gives reasonable
argumentation of the fact that the bifurcations cascade can be expressed in
terms of length scales.
Furthermore, the study analyzes the characteristic length scales at the
transition through global properties of the bifurcations, and estimates the
order of magnitude of the
critical Taylor--scale Reynolds number and the number of bifurcations at the
onset of turbulence.
\end{abstract}

\begin{keyword}

Energy cascade, Bifurcations, Fixed points,  Lyapunov theory.
\end{keyword}

\end{frontmatter}

\newcommand{\no}{\noindent}
\newcommand{\be}{\begin{equation}}
\newcommand{\ee}{\end{equation}}
\newcommand{\bea}{\begin{eqnarray}}
\newcommand{\eea}{\end{eqnarray}}
\newcommand{\bc}{\begin{center}}
\newcommand{\ec}{\end{center}}

\newcommand{\calr}{{\cal R}}
\newcommand{\calv}{{\cal V}}

\newcommand{\bff}{\mbox{\boldmath $f$}}
\newcommand{\bfg}{\mbox{\boldmath $g$}}
\newcommand{\bfh}{\mbox{\boldmath $h$}}
\newcommand{\bfi}{\mbox{\boldmath $i$}}
\newcommand{\bfm}{\mbox{\boldmath $m$}}
\newcommand{\bfp}{\mbox{\boldmath $p$}}
\newcommand{\bfr}{\mbox{\boldmath $r$}}
\newcommand{\bfu}{\mbox{\boldmath $u$}}
\newcommand{\bfv}{\mbox{\boldmath $v$}}
\newcommand{\bfx}{\mbox{\boldmath $x$}}
\newcommand{\bfy}{\mbox{\boldmath $y$}}
\newcommand{\bfw}{\mbox{\boldmath $w$}}
\newcommand{\bfk}{\mbox{\boldmath $\kappa$}}

\newcommand{\bfA}{\mbox{\boldmath $A$}}
\newcommand{\bfD}{\mbox{\boldmath $D$}}
\newcommand{\bfI}{\mbox{\boldmath $I$}}
\newcommand{\bfL}{\mbox{\boldmath $L$}}
\newcommand{\bfM}{\mbox{\boldmath $M$}}
\newcommand{\bfS}{\mbox{\boldmath $S$}}
\newcommand{\bfT}{\mbox{\boldmath $T$}}
\newcommand{\bfU}{\mbox{\boldmath $U$}}
\newcommand{\bfX}{\mbox{\boldmath $X$}}
\newcommand{\bfY}{\mbox{\boldmath $Y$}}
\newcommand{\bfK}{\mbox{\boldmath $K$}}

\newcommand{\bfrho}{\mbox{\boldmath $\rho$}}
\newcommand{\bfchi}{\mbox{\boldmath $\chi$}}
\newcommand{\bfphi}{\mbox{\boldmath $\phi$}}
\newcommand{\bfPhi}{\mbox{\boldmath $\Phi$}}
\newcommand{\bflambda}{\mbox{\boldmath $\lambda$}}
\newcommand{\bfxi}{\mbox{\boldmath $\xi$}}
\newcommand{\bfLambda}{\mbox{\boldmath $\Lambda$}}
\newcommand{\bfPsi}{\mbox{\boldmath $\Psi$}}
\newcommand{\bfomega}{\mbox{\boldmath $\omega$}}
\newcommand{\bfOmega}{\mbox{\boldmath $\Omega$}}
\newcommand{\bfeps}{\mbox{\boldmath $\varepsilon$}}
\newcommand{\bfepsn}{\mbox{\boldmath $\epsilon$}}
\newcommand{\bfzeta}{\mbox{\boldmath $\zeta$}}
\newcommand{\bfkappa}{\mbox{\boldmath $\kappa$}}
\newcommand{\bfsigma}{\mbox{\boldmath $\sigma$}}
\newcommand{\itPsi}{\mbox{\it $\Psi$}}
\newcommand{\itPhi}{\mbox{\it $\Phi$}}

\newcommand{\bint}{\mbox{ \int{a}{b}} }
\newcommand{\ds}{\displaystyle}
\newcommand{\Sum}{\Large \sum}


\section{Introduction \label{intro}}

The main aim of this work is to analyze the turbulent mechanism of energy
cascade by means of the bifurcations properties of the Navier--Stokes equations.
This analysis is accomplished in the case of incompressible fluid in an infinite domain.
Moreover, through this analysis, we want to corroborate the basic hypotheses of
previous 
works \cite{deDivitiis_1, deDivitiis_4} where the finite--scale Lyapunov
theory is used to describe the homogeneous isotropic turbulence.
There, the theory, which leads to the closure of von K\'arm\'an--Howarth and
Corrsin  equations \cite{Karman38, Corrsin_1}, is 
also based on the assumption that the bifurcations cascade law can be expressed
in terms of length scales, and on the hypothesis that the relative kinematics
between two contiguous particles is much faster than the fluid state variables.
This latter is justified by the fact that in turbulence the kinematics of fluid
deformation exhibits chaotic behavior and huge mixing 
\cite{Ottino89, Ottino90}, and allows to express velocity and temperature fluctuations with
Navier--Stokes and temperature equations, through the local fluid deformation
\cite{deDivitiis_1, deDivitiis_4}.

The study first introduces the bifurcations of the Navier--Stokes equations
(NS--bifurcations), in line with the classical theory of differential equations
\cite{Ruelle71, Eckmann81, Guckenheimer90} and shows that the local fluid deformation
$\partial{\bf x}/\partial{\bf x}_0$ can be much more rapid than the fluid state
variables.
The phenomenon of energy cascade is then studied through a statistical property
of the
Navier--Stokes equations in regimes of fully developed chaos. This property,
which represents an important element of this work, is obtained through basic
characteristics of the bifurcations in developed chaos, 
and gives the link between NS--bifurcations and energy cascade mechanism.

Next, to found the link between length scales and NS--bifurcations, the fixed
points of the velocity field and the corresponding bifurcations
(u--bifurcations) are properly defined.
According to this definition, these u--bifurcations are shown to be
non--material moving points which represent the trace of the NS--bifurcations in
the fluid domain. Such spatial representation justifies the fact that the
bifurcations cascade can be expressed in terms of length scales.
This representation, together to general properties of the route toward the
chaos and to the fractal characteristics of the bifurcations in developed
turbulence, allows to analytically express one  reasonable bifurcations cascade
law in terms of length scales. 
{\color{black} Of course, the present analysis can not give the physical meaning of
the characteristic lengths, because it studies the Navier--Stokes equations in an infinite region. 
For what concerns the physical meaning of these scales, one can follow the procedure proposed 
by Ran Zheng \cite{RAN_Zheng11}, where, starting from the previous results of Sedov \cite{Sedov44}, 
the question is addressed with an opportune analysis of the closed von K\'arm\'an--Howarth equation 
and assuming the hypotheses of similarity and self--preservation. The closure of such equation and the self--similarity allow to obtain the analytical structure of the solutions compatible with the boundary conditions, where both second and triple pair velocity correlations are in terms of one reduced dimensionless variable. 
From this analysis, the author \cite{RAN_Zheng11} obtains one evolution equation for the length scale 
(for further details see also \cite{RAN_Zheng09}) which, among the other things, admits two kinds of solution. This leads to a natural definition of the scales cascade process, where this latter is shown to be a multistage process.}

\bigskip

Through the elements of the present work, we furnish plausible argumentations that the
NS--bifurcations are responsible for the main properties of turbulence, such as
the chaotic fluid motion, the energy cascade, the continuous distribution of the
length scales.

Moreover, a relationship between the order of magnitude of $R_{\lambda}^*$ and
$N$ is found,
where  $R_{\lambda}^*$ and $N$ are, respectively, the critical Taylor--scale
Reynolds number 
and number of bifurcations at the transition.
This estimation, based on adequate hypotheses about the length scales, gives
$N=3$ and  
$R_{\lambda}^*$ = 4$\div$14, in agreement with the several theoretical and
experimental sources of the literature.
Next, $R_{\lambda}^*$ is also estimated beginning from the fully developed
isotropic turbulence using the closed von K\'arm\'an--Howarth equation 
proposed by the author in Ref. \cite{deDivitiis_1}.
The two procedures give results in agreement with each other.

\bigskip

\section{Background: Bifurcations of Navier--Stokes equations
\label{Background}}

This section recalls some of the elements of the Navier--Stokes equations 
and the general properties of the bifurcations 
which are necessary for the purposes of the present analysis. 

The dimensionless Navier--Stokes equations are
\bea
\begin{array}{l@{\hspace{-0.cm}}l}
\ds \nabla \cdot {\bf u} =0, \\\\
\ds \frac{\partial {\bf u}}{\partial t} =
-{\bf u} \cdot \nabla {\bf u}  - \nabla p + \frac{1}{Re} \nabla^2 {\bf u}
\end{array}
\label{NS_eq}
\eea
where $Re = U L/\nu$ is the Reynolds number, $\bf u$=${\bf u}({\bf x}, t)$
and $p$=$p({\bf x}, t)$ are dimensionless velocity and pressure, 
whereas $U$ and $L$ are the reference velocity and length.
For sake of convenience, the momentum Navier--Stokes equations 
are formally written by eliminating the pressure field in Eqs. (\ref{NS_eq}) 
through the continuity equation
\bea
 \dot{\bf u} = {\bf N}({\bf u} ; Re) \equiv
 {\bf N}_0({\bf u}) + \frac{1}{Re} {\bf L} {\bf u}
\label{NS_sint1}
\eea
where $\dot{\bf u}$ is the Eulerian time derivatives of the velocity field,
\bea
\ds {\bf N}: \{ {\bf u} \}
 \rightarrow 
\ds  \left\lbrace  \frac{\partial {\bf u}}{\partial t} \right\rbrace 
\label{op_NS}
\eea 
is the nonlinear operator representing the R.H.S. of the momentum Navier--Stokes
equations, and $ \{ {\bf u} \}$ and 
$\left\lbrace {\partial {\bf u}}/{\partial t} \right\rbrace$
are the sets of $\bf u$ and ${\partial {\bf u}}/{\partial t}$,
respectively.
In Eq. (\ref{NS_sint1}), ${\bf N}_0({\bf u})$ is the nonlinear quadratic
operator representing the inertia and pressure forces, whereas the linear
operator ${\bf L} {\bf u}$ gives the viscosity term.

In the case of homogeneous fluid in infinite domain, if ${\bf u}({\bf x}, t)$ is
a solution of Eq. (\ref{NS_sint1}), then ${\bf u}({\bf x} + {\bf h}, t)$ satisfies Eq.
(\ref{NS_sint1}), where ${\bf h}$ is an arbitrary displacement, i.e.
\bea
\ds \dot{\bf u}({\bf x}, t) = {\bf N}({\bf u}({\bf x}, t) ; Re) \ \ \Rightarrow
\ \
\ds \dot{\bf u}({\bf x} + {\bf h}, t) = {\bf N}({\bf u}({\bf x} + {\bf h}, t) ;
Re), \ \forall  {\bf h}
\label{homo_inf}
\eea

{\color{black} Observe that, ${\bf N}_0({\bf u})$ incorporates the integral 
non--linear operator which expresses the pressure gradient through the velocity 
field in the entire fluid domain, where
\bea
\ds \nabla^2 p = - 
 \frac{\partial^2 u_i u_j}{\partial x_i \partial x_j} 
\nonumber
\eea
This produces the non--local nonlinear influence of the velocity field on 
$\partial {\bf u}/ \partial t$ in the entire fluid domain
\cite{Tsinober2009}, thus the Navier--Stokes equations are reduced to be an integro--differential equation formally expressed by Eq. (\ref{NS_sint1}) in the symbolic form of operators. 
Although there is no explicit methods to formulate the bifurcation analysis for integro--differential equations, in line with Ruelle and Takens \cite{Ruelle71}, we suppose that $ \{ {\bf u} \}$, which is an infinite dimensional vector space, can be replaced by a finite-dimensional manifold, thus
the Navier--Stokes equations in the form (\ref{NS_sint1}) correspond to the equation studied by Ruelle and Takens. Therefore,  Eq. (\ref{NS_sint1}) is analyzed through the classical bifurcation theory of differential equations \cite{Ruelle71, Eckmann81}, and the results here obtained can be considered to be valid in the limits of the formulation proposed by Ruelle and Takens in Ref. \cite{Ruelle71}.}

Now, to define the bifurcations of the Navier--Stokes equations, 
consider the steady solutions of Eq. (\ref{NS_sint1})
\bea
{\bf N} ({\bf u}; Re) = 0
\label{NS_sint2}
\eea
If $Re=Re_0$ is properly small, the unique steady solution 
${\bf u} (Re_0) = {\bf u}({\bf x}; Re_0)$ is calculated by inversion of  Eq.
(\ref{NS_sint2}),
and for $Re>Re_0$, the steady solutions ${\bf u}(Re)$ 
can be obtained starting from 
${\bf u}(Re_0)$, by applying the implicit function theorem to Eq.
(\ref{NS_sint2})
\bea
\ds {\bf u}(Re) = 
\ds {\bf u}(Re_0) -
\ds  \int_{Re_0}^{Re} {\nabla_{\bf u} {\bf N}}^{-1} \ \frac{\partial {\bf
N}}{\partial Re} \ dRe
\label{bif_map_v}
\eea 
where 
$
\nabla_{\bf u} {\bf N} 
\equiv 
{\partial {\bf N}({\bf u}; Re)}/{\partial {\bf u}}
$ is
the Jacobian of $\bf N$ with respect to ${\bf u}$.
The velocity field ${\bf u}(Re)$ can be determined with Eq. (\ref{bif_map_v}) as
long as $\nabla_{\bf u} {\bf N}$ is nonsingular, i.e. when the determinant 
$\det (\nabla_{\bf u} {\bf N}) \ne 0$.

The bifurcations of the Navier--Stokes equations occur when ${\nabla_{\bf u}
{\bf N}}$ exhibits at least an eigenvalue with zero real part
(NS--bifurcations). 
There, $\det (\nabla_{\bf u} {\bf N}) =0$ thus, following Eq. (\ref{bif_map_v}),
${\bf u}(Re)$ can degenerate in two or more solutions.
As the consequence of the structure of Eq. (\ref{NS_sint1}),
we have the following route toward the chaos:
For small $Re$, the viscosity forces are stronger than the inertia ones and 
$\bf N$ behaves like a linear operator with $\det (\nabla_{\bf u} {\bf N}) \ne
0$.
When the Reynolds number increases, as long as $\nabla_{\bf u} {\bf N}$ is
nonsingular, $\partial {\bf u} / \partial Re$ exhibits smooth variations, 
whereas at a certain $Re$, this Jacobian becomes singular and $\partial {\bf u}
/ \partial Re$ appears to be discontinuous with respect to $Re$.
Of course, the route toward the turbulence can be of different kinds,
such as, for example, those described in Refs. \cite{Ruelle71, Feigenbaum78, Pomeau80}.
In general, the chaotic motion is observed when the number of encountered
bifurcations is about greater than three.
\begin{figure}[h]
\centering
\vspace{-0.mm}
\hspace{-0.mm}
\includegraphics[width=0.45 \textwidth]{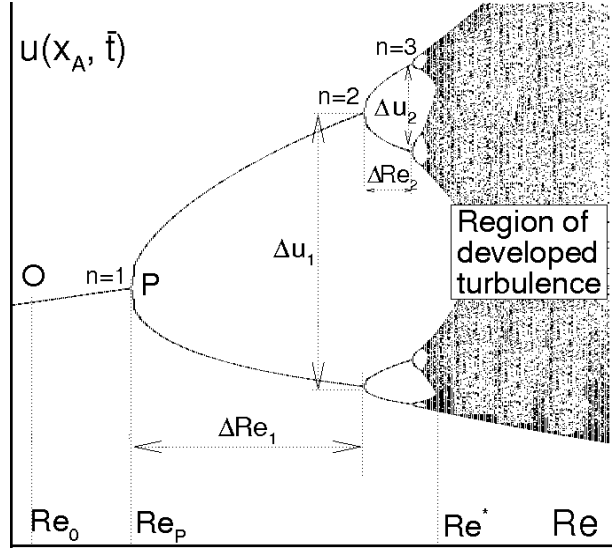}
\caption{Qualitative scheme of NS--bifurcations.}
\label{figura_1}
\end{figure}
Figure \ref{figura_1} reports a scheme of the bifurcations tree. {\color{black} This and the following schemes of bifurcations are purely qualitative schemes the purpose of which is only to show the necessary elements for this analysis}. In the figure one component of ${\bf u}({\bf x}_A)$ is shown in function of $Re$, where ${\bf x}_A$ ia an assigned position.
Starting from $Re_0$, the diagram is regular, until $Re_P$, where the first
bifurcation determines two branches.
Increasing again $Re$, other bifurcations occur. In the figure,
$\Delta u$ denotes the distance between two branches which born from the same
bifurcation, 
$\Delta Re$ represents the distance between two successive bifurcations and $n$
is the
number of the encountered bifurcations starting from $Re_0$. 

If the Reynolds number does not exceed its critical value value, say $Re^*$, 
the velocity fields satisfying Eq. (\ref{bif_map_v}) are limited in number, thus
also $n$ is moderate, 
and the branches corresponds to the intermediate stages of the route toward the
chaos.

Conversely, when $Re > Re^*$, we have the region of developed turbulence. There,
$\lambda_{NS}>$0,
where $\lambda_{NS}$ is the maximal Lyapunov exponent of the Navier--Stokes
equations, 
formally calculated as
\bea
\begin{array}{l@{\hspace{-0.cm}}l}
\ds \lambda_{NS} = \lim_{T \rightarrow \infty} \frac{1}{T} \int_0^T 
\frac{{\bfy} \cdot \nabla_{\bf u} {\bf N} {\bfy} }{{\bfy} \cdot{\bfy}} \ dt,
\\\\
\ds \dot{\bfy} = \nabla_{\bf u} {\bf N}({\bf u}; Re) {\bfy}, \\\\
\ds \dot{\bf u} = {\bf N} ({\bf u}; Re),
\end{array}
\eea
and $\bfy$ is the Lyapunov vector of the Navier--Stokes equations. Accordingly,
$Re^*$ depends on ${\bf u}$, and is defined as the minimum value of the Reynolds
number at which $\lambda_{NS} \geq$ 0.
For $Re>Re^*$, the diverse velocity fields satisfying Eq. (\ref{bif_map_v}),
determine an extended complex geometry made by several points whose minimum
distance is very small. There, the equation ${\bf N}({\bf u}, Re) =$0 is satisfied in a
huge number of points of $\left\lbrace \bf u \right\rbrace $  which are very close with each
other, whereas ${\bf N} ( {\bf u}, Re) \ne$0  elsewhere.
This corresponds to a situation in which $\bf u$, $\bf N$ and $\nabla_{\bf u}
{\bf N}$ exhibit abrupt variations. 

It is worth to remark two general properties of the bifurcation diagrams.
1) The first property regards the overall dimension along $u$ of the
bifurcations tree: following such property, this dimension varies quite
regularly with respect to $Re$ also through the transition, 
and can be represented by one proper smooth rising function of $Re$ of class $C^0(Re)$.
2) The other property pertains the beginning of chaos: the region of developed
turbulence is bounded at the onset of turbulence by bifurcations lines which
separate the region of the developed chaos by the remaining zone. These lines
form bifurcation tongues, regions of developed chaos, whose local extension
along $\Delta u$ increases with $Re$ until to overlap with each other.

The bifurcations are also responsible for the sharp variations of 
the characteristic length scales of the velocity fields.
For $Re<<<Re^*$, the solutions of Eq. (\ref{NS_sint2}) 
can be expressed by Fourier series of a given characteristic length scale. 
When $Re$ steadily rises, each encountered bifurcation introduces new solutions
whose characteristic scales are independent from the previous ones until to reach 
the edge of turbulence where $Re \lesssim Re^*$. These are independent scales discretely distributed.
Thereafter, for $Re\gtrsim Re^*$, such discrete distribution disappears, and the scales seem to 
be continuously distributed.

\bigskip

\section{Bifurcation of unsteady solutions and divergence of phase trajectories}

In the case of unsteady flow, the bifurcations are responsible for multiple
unsteady velocity fields.
In fact, during the motion, multiple solutions 
$\hat {\bf u}$ can be determined, at each instant, through inversion of Eq.
(\ref{NS_sint1})
\bea
\begin{array}{l@{\hspace{-0.cm}}l}
\dot{\bf u} = {\bf N}({\bf u}; Re)  \\\\
\hat {\bf u} (Re) = {\bf N}^{-1}(\dot{\bf u}; Re) 
\end{array}
\label{inv}
\eea
If $Re << Re^*$, $\bf N$ behaves like a linear operator, and 
Eq. (\ref{inv}) gives $\hat {\bf u} \equiv {\bf u} ({\bf x}, t)$ as unique
solution, 
whereas if $Re$ is properly high, ${\bf N}^{-1}$ is a multivalued operator and
Eq. (\ref{inv}) determines several velocity fields $\hat {\bf u}$. 
That is, the current velocity field ${\bf u} ({\bf x}, t)$ 
corresponds to several other solutions $\hat {\bf u} ({\bf x}, t; Re)$ which
give the same field $\dot{\bf u} ({\bf x}, t)$. For $Re > Re^*$ a huge number of
these solutions are unstable, thus the bifurcations determine a situation
where ${\bf u} ({\bf x}, t)$ tends to sweep the entire velocity field set, 
accordingly the motion is expected to be chaotic with a high level of mixing.

\bigskip

When $Re$ is given, a single NS--bifurcation corresponds
to several trajectories bifurcations in the hodograph space and to a
growth of
the velocity gradient $\nabla_{\bf x} {\bf u}$.
To show this, consider now the two velocity fields ${\bf u}={\bf u} ({\bf x},
t)$ and 
${\bf u}' ({\bf x}, t)= {\bf u} ({\bf x}+{\bf r}, t)$, where $\bf r$, and 
${\bfxi}$ = ${\bf u}'$ - ${\bf u}$. 
\begin{figure}[h]
\centering
\vspace{-0.mm}
\hspace{-0.mm}
\includegraphics[width=0.4 \textwidth]{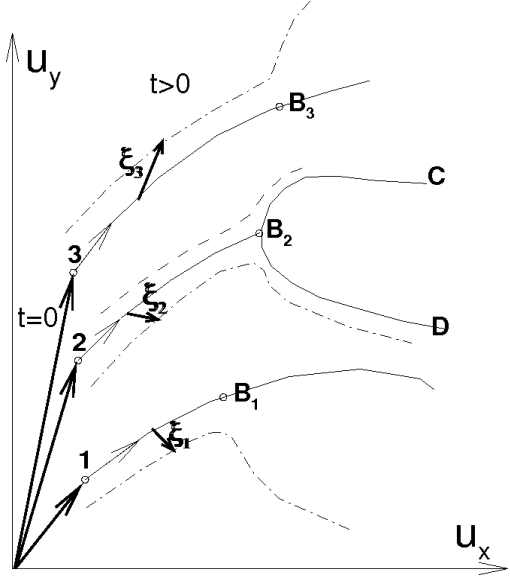}
\caption{Qualitative scheme of phase trajectories in the hodograph plane of
three different points
of the space.}
\label{figura_1bis}
\end{figure}
Thanks to the property (\ref{homo_inf}) (homogeneous fluid in infinite space),
$\bf u$ and ${\bf u}'$ are both solutions of Eq. (\ref{NS_sint1}), thus the
evolution equations of ${\bfxi}$ coincide with those of perturbation of the velocity field, 
and if $r = \vert {\bf r} \vert$ is quite small, these equations are 
\bea
\begin{array}{l@{\hspace{-0.cm}}l}
\dot{\bfxi} = \nabla_{\bf u} {\bf N} \ {\bfxi}, \\\\
\dot{\bf u} = {\bf N}({\bf u}; Re) 
\end{array}
\label{NSe}
\label{NSxi}
\eea 
Hence, the classical theory of the differential equations can be applied 
to Eq. (\ref{NSe}).
For sake of simplicity, we suppose that at the onset of the motion 
all the eigenvalues of $\nabla_{\bf u} {\bf N}({\bf u}, Re)$ exhibit negative
real part, and that
the NS--bifurcation happens for $t=t^*>0$. There, at least an eigenvalue of
$\nabla_{\bf u} {\bf N}({\bf u}, Re)$ crosses the imaginary axis, and the phase
trajectories, initially contiguous, thereafter diverge.
Figure \ref{figura_1bis} shows three pairs of trajectories in the same hodograph
plane 
($u_x$, $u_y$), each representing the velocity components in the pairs of points
(${\bf x}_1$, ${\bf x}_1 + {\bf r}$), (${\bf x}_2$, ${\bf x}_2 + {\bf r}$) 
and (${\bf x}_3$, ${\bf x}_3 + {\bf r}$), 
where the arrows denote increasing time.
Continuous and dashed lines represent the velocities calculated in  
${\bf x}_1$, ${\bf x}_2$, ${\bf x}_3$ and 
${\bf x}_1 + {\bf r}$, ${\bf x}_2 + {\bf r}$, ${\bf x}_3 + {\bf r}$, 
respectively, whereas the points ${\bf B}_1$, ${\bf B}_2$, ${\bf B}_3$ give the
velocities
at $t=t^*$, thus these are the image of the NS--bifurcation in the hodograph
plane. 
After the NS--bifurcation ${\bfxi_1}(t)$, ${\bfxi_2}(t)$ and ${\bfxi_3}(t)$
diverge, 
and this means that a bifurcation causes the lost of informations with respect 
to the initial condition ${\bfxi}_1(0)$, ${\bfxi_2}(0)$ and ${\bfxi_3}(0)$.
In particular, for what concerns the single trajectory $\bf 2$--${\bf B}_2$, 
after ${\bf B}_2$ it degenerates in the two branches ${\bf B}_2-{\bf C}$ and
${\bf B}_2-{\bf D}$
which represent two possible phase trajectories, thus $\partial{\bf u}/\partial
t$ =0
in ${\bf B}_2$.
After ${\bf B}_2$, Eq. (\ref{NSe}) does not indicate which of the branches the
fluid will choose,
thus very small variations on the initial condition or little perturbations, are
of paramount importance for the choice of the branch that the fluid  will follow
\cite{Prigogine94}.

\bigskip

\section{Property of the local fluid deformation in developed turbulence \label{fluid deformation}}

This section presents reasonable argumentations that, in turbulence,  
the local fluid deformation
\bea
\ds \frac{\partial {\bf x}}{\partial {\bf x}_0}
\eea 
can be much more rapid than the fluid state variables, where ${\bf x}_0$ and
${\bf x}$ are material
coordinates, and the function $\chi : {\bf x}_0 \rightarrow {\bf x}$ gives the
current position ${\bf x}$ of a fluid particle located at the referential
position ${\bf x}_0$ at $t=t_0$.
To show this, observe that, in Eq. (\ref{NSxi}), $\bfxi$ corresponds to
variations of the 
velocity gradient $\nabla_{\bf x} {\bf u}$ which thus changes according to
\bea
{\nabla_{\bf x} {\bf u}} ({\bf x}, t) = 
\exp{ \left( \int_0^t \nabla_{\bf u} {\bf N} dt \right) }  
{\nabla_{\bf x} {\bf u}} ({\bf x}, 0)
\label{initG0}
\eea
where the exponential denotes the series expansion of operators 
\bea
\begin{array}{l@{\hspace{-0.cm}}l}
\ds \exp{ \left( \int_0^t \nabla_{\bf u} {\bf N} dt \right) }  
=  {\bf I} + \int_0^t {\nabla_{\bf u} {\bf N}} \ dt  + ...  
\end{array}
\label{initG}
\eea
${\nabla_{\bf x} {\bf u}} ({\bf x}, 0)$
is the initial condition, and $\bf I$ is the identity map.
The bifurcations determine abrupt changing in ${\nabla_{\bf u} {\bf N}}$ which
in turn produces 
important increments of the velocity gradient according to Eq. (\ref{initG0}).
Although such variations are very significant and $\lambda_{NS}>$0, due to the
viscosity, 
$\nabla_{\bf x} {\bf u}$ is however a function of slow growth of $t \in(t_0,
\infty)$.

On the contrary, $\partial {\bf x}/ \partial {\bf x}_0$ is not bounded by
the dissipation effects. 
This deformation, related to the relative kinematics between two contiguous
particles,
is represented by the infinitesimal separation vector 
$d {\bf x}$ between the particles, which varies according to
\bea
 {d \dot {\bf x}} = \nabla_{\bf x} {\bf u} \ d {\bf x}
\label{kin 0}
\eea
The Lyapunov analysis of this equation gives the local deformation in terms of
the
maximal Lyapunov exponent $\Lambda = \max (\Lambda_1, \Lambda_2, \Lambda_3)$
associated to Eq. (\ref{kin 0})
\bea
\ds \frac{\partial {\bf x}}{\partial {\bf x}_0} \approx e^{\Lambda(t-t_0)}
\label{strain}
\eea
where $\Lambda_1$, $\Lambda_2$ and $\Lambda_3$ are the Lyapunov exponents of Eq.
(\ref{kin 0}).
Due to the fluid incompressibility $\Lambda>0$, and as $\vert \vert {\nabla_{\bf
x} {\bf u}} \vert \vert >>0$, the exponent $\Lambda \approx \vert \vert
{\nabla_{\bf x} {\bf u}} \vert \vert $
is expected to be high, thus according to Eq. (\ref{strain}) 
$\partial {\bf x}/\partial {\bf x}_0$ can be much faster than ${\nabla_{\bf x}
{\bf u}}$ and $\bf u$.
Therefore, as the local fluid deformation is not bounded by the viscosity,
$\partial {\bf x}/ \partial {\bf x}_0$ is represented by functions of
exponential growth, whereas the fluid state variables are functions of slow
growth of $t \in(t_0, \infty)$.

\bigskip

{\bf Remark}. This property can have implications for what concerns the basic
formulation for deriving the Navier--Stokes equations.
In fact, momentum and continuity equations are derived from an integral
formulation of balance equations by means of the Green theorem, and this latter
can be applied to regions which exhibit smooth boundaries during the motion
\cite{Truesdell77}.
Now, if $\partial {\bf x}/ \partial {\bf x}_0$ is much more rapid than $\bf u$
and exhibits abrupt spatial variations, the boundaries of fluid region become
irregular in very short times, and this implies that  momentum and continuity 
equations could require the consideration of very small scales and times for
describing the fluid motion \cite{Truesdell77} .

\bigskip

\section{Relationship bifurcations--energy cascade in fully developed turbulence
 \label{energy cascade}}

\begin{figure}[h]
\centering
\vspace{-0.mm}
\hspace{-0.mm}
\includegraphics[width=0.5500\textwidth]{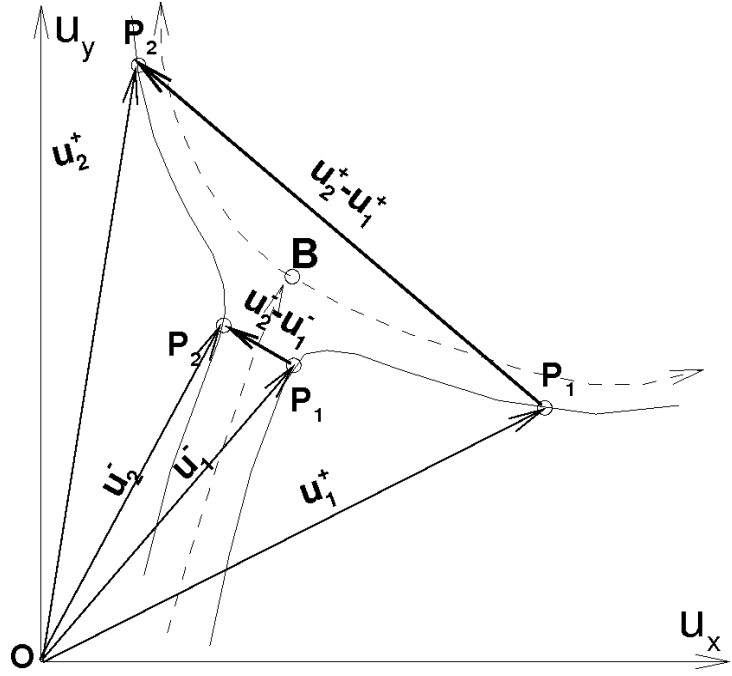}
\caption{Scheme of the velocities variations near a bifurcation in the hodograph
plane.}
\label{figura_bif}
\end{figure}

The purpose of this section is to show the link between the bifurcations
and the phenomenon of turbulent energy cascade.
To study this, a simple statistical property of the Navier--Stokes equations in
the regime of  fully developed chaos is proposed.
This property, arising from basic elements of the bifurcations, is here applied
to the Navier--Stokes equations in the form (\ref{NS_sint1}).
To this end, consider now Fig. \ref{figura_bif}, where a scheme of two
contiguous phase trajectories in the hodograph plane is shown in proximity of
the trajectory bifurcation
$\bf B$.
These trajectories correspond to velocity variations in two assigned
points 
${\bf x}_1$ and  ${\bf x}_2 = {\bf x}_1 + {\bf r}$, 
where $r = \vert {\bf r} \vert > 0$ is arbitrarily small.
The figure shows the velocity arrows $P_1$ and $P_2$ which describe the two
phase trajectories, initially close with each other, that thereafter diverge
because of the bifurcation.
Let $t^-$ and $t^+$ instants for which both $P_1$ and $P_2$ approach to $\bf B$
and move away from it, respectively.
In developed turbulence, $\lambda_{NS}>$0, and the phase trajectories diverge,
i.e. 
\bea
\ds \vert  {\bf u}_2^+ - {\bf u}_1^+  \vert >>
\ds \vert  {\bf u}_2^- - {\bf u}_1^-  \vert
\label{cond_0}
\eea
moreover thanks to the numerous bifurcations $\det ( \nabla_{\bf u} {\bf N})
\approx 0$, therefore we expect that 
\bea
\vert {\bf u}_1^+ \vert \approx \vert {\bf u}_1^- \vert, \ \
\vert {\bf u}_2^+ \vert \approx \vert {\bf u}_2^- \vert
\label{cond_b}
\eea 
The inequality (\ref{cond_0}) and Eq. (\ref{cond_b})  imply that
\bea
\ds {\bf u}_1^+ \cdot \left(  {\bf u}_2^+ - {\bf u}_1^+ \right) <<
\ds {\bf u}_1^- \cdot \left(  {\bf u}_2^- - {\bf u}_1^- \right)
\label{cond_1}
\eea
The condition (\ref{cond_1}) is frequentely satisfied in the chaotic regime, 
whereas the opposite inequality is possible but not probable. 
Hence, form the statistical point of view, it is reasonable that 
\bea
\ds \frac{\partial}{\partial t}
\ds  \left\langle  {\bf u} \cdot \left(  {\bf u}' - {\bf u} \right)
\right\rangle
 \leq 0,
\ \ \ \forall r  \ \mbox{small}
\label{cond_1a}
\eea
where ${\bf u} = {\bf u}_1$, ${\bf u}' = {\bf u}_2$, and $\langle . \rangle$
denotes the average over the velocity ensemble.
Moreover, for relatively high values of $r$, due to the numerous trajectory 
bifurcations in between ${\bf x}_1$ and ${\bf x}_2$,
the inequality (\ref{cond_1}) will be satisfied in average, thus we assume that 
\bea
\ds \int_V \frac{\partial}{\partial t}
\ds  \left\langle  {\bf u} \cdot \left(  {\bf u}' - {\bf u} \right)
\right\rangle \ dV'  \leq 0,
\ \ \ \forall \ V, 
\label{cond_2}
\eea
in which $dV' = dr_x dr_y dr_z$ is the elemental volume with 
$d{\bf r} = (dr_x, dr_y, dr_z)$,
and $\int_V \ dV' = 4/3 \pi r^3$.
Taking into account Eq. (\ref{NS_sint1}), Eqs. (\ref{cond_2}) and
(\ref{cond_1a}) 
are both summarized by the following condition
\bea
\begin{array}{l@{\hspace{-0.cm}}l}
\ds \int_V  \left\langle {\bf N} \cdot \Delta {\bf u}  +
\ds {\bf u} \cdot \Delta {\bf N}  \right\rangle   \ dV' \leq 0,
\ \ \ \forall \ V
 \end{array}
\label{cond_2a}
\eea
where ${\bf N} = {\bf N}_1$, $\Delta {\bf N} = {\bf N}' -{\bf N}_1$, and 
$\Delta {\bf u} = {\bf u}' -{\bf u}$. 
Observe that Eq. (\ref{cond_2a}) has been 
obtained from Eq. (\ref{NS_sint1}), for arbitrary $Re >Re^*$.
Due to this arbitrarily and considering that the bifurcations are caused by the
nonlinear terms of the Navier--Stokes equations,  we obtain 
\bea
\begin{array}{l@{\hspace{-0.cm}}l}
\ds \int_V \left\langle {\bf N}_0 \cdot \Delta {\bf u}  +
\ds {\bf u} \cdot \Delta {\bf N}_0  \right\rangle  \ dV' \leq 0,
\ \ \ \forall \   V  
\end{array}
\label{cond_3}
\eea
Equation (\ref{cond_3}) represents the proposed relationship, which
expresses the influence of the bifurcations on the fluid motion.

At this stage of the analysis, we argue that the
bifurcations determine the transfer of kinetic energy from large to small
scales.
To demonstrate this, we will show, starting from Eq. (\ref{cond_3}), that
\bea
\begin{array}{l@{\hspace{-0.cm}}l}
\ds H_3(r) < 0 \ \  \forall r \ge 0, \\\\
\ds \lim_{r \rightarrow \infty} H_3(r) =0
\end{array}
\eea
in the case of homogeneous isotropic turbulence, where $H_3(r)$ is the third
dimensionless statistical
moment (skewness) of the longitudinal component of ${\bf u}'-{\bf u}$.
This is shown by means of the
evolution equation of the velocity correlation.
This equation is obtained through the Navier--Stokes equations written in two
points $\bf x$ and ${\bf x'}=$ ${\bf x} + {\bf r}$, taking into account that, in
such condition 
$\langle {\bf N}_0 {\bf u} \rangle \equiv 0$ \cite{Karman38}
\bea
\begin{array}{l@{\hspace{-0.cm}}l}
\ds \frac{\partial}{\partial t} \left\langle  {\bf u} \cdot {\bf u}'
\right\rangle  = 
 Re^{-1} \left(  2  \left\langle  {\bf u} \cdot {\bf L} {\bf u} \right\rangle 
+ \left\langle {\bf L} {\bf u} \cdot \Delta {\bf u} + 
{\bf L} \Delta {\bf u} \cdot {\bf u} \right\rangle \right)  \\\\
+ \left\langle {\bf N}_0 \cdot \Delta {\bf u} + \Delta {\bf N}_0 \cdot {\bf u}
\right\rangle 
\end{array}
\label{corr}
\eea
The first integral of Eq. (\ref{corr}) is the evolution equation of 
the longitudinal velocity correlation function \cite{Karman38}.
First and second terms at the R.H.S. of Eq. (\ref{corr}) give respectively,
the rate of kinetic energy and the spatial variations of the velocity
correlation due to the viscosity, whereas the third one, arising from the
inertia forces, is responsible for the mechanism of energy cascade and
identifies the term with $H_3(r)$ \cite{Karman38}
\bea
\ds \left\langle 
{\bf N}_0 \cdot \Delta {\bf u} + \Delta {\bf N}_0 \cdot {\bf u} 
\right\rangle
= \nabla \cdot \left\langle ({\bf u}\cdot {\bf u}' ) ({\bf u}-{\bf u}')
\right\rangle 
\eea
where
\bea
\ds \left\langle {\bf N}_0 \cdot \Delta {\bf u}  +
\ds {\bf u} \cdot \Delta {\bf N}_0  \right\rangle
=0 \ \ \ \mbox{for} \ r=0,
\label{K0=0} 
\eea
and
\bea
\lim_{r \rightarrow \infty} 
\ds \left\langle {\bf N}_0 \cdot \Delta {\bf u}  +
\ds {\bf u} \cdot \Delta {\bf N}_0  \right\rangle
=0,
\label{K00=0} 
\eea
Following Eq. (\ref{K0=0}) the bifurcations do not modify the average kinetic
energy rate, 
thus they are only responsible for the energy cascade, 
whereas Eq. (\ref{K00=0}) states that their effect vanishes
for $r \rightarrow \infty$.

Now, in line with Ref. \cite{Karman38}
\bea
\begin{array}{l@{\hspace{-0.cm}}l}
\ds \frac{1}{r^2} \frac{\partial}{\partial r} \left( r^3 K(r) \right) \equiv
\ds \nabla \cdot \left\langle ({\bf u}\cdot {\bf u}' ) ({\bf u}-{\bf u}')
\right\rangle
\end{array}
\label{ka}
\eea
where $K(r)$ is an even function of $r$ directly related to the longitudinal
triple 
correlation function $k(r) = \langle u_r^2 u_r' \rangle/ u^3$, according to
\bea
\begin{array}{l@{\hspace{-0.cm}}l}
\ds \frac{1}{r^4} \frac{\partial}{\partial r} (r^4 k(r)) = \frac{K(r)}{u^3}
\end{array}
\label{kb}
\eea
where $u = \sqrt{\langle u_r^2 \rangle} \equiv \sqrt{\langle {\bf u}\cdot{\bf
u}\rangle/3}$, 
$u_r = {\bf u} \cdot {\bf r}/r$ and $k$ is an odd function of $r$ which vanishes
for 
$r \rightarrow \infty$ and such that $k=O(r^3)$ near the origin
\cite{Karman38}.
Now, integrating Eq. (\ref{ka}) with respect to the volume $V$ and taking into
account 
Eq. (\ref{cond_3}) and that $K(0)=0$, we have $K(r) < 0 \ \forall r >0$, where
due to
isotropy,   $dV' = dr_x dr_y dr_z = 4 \pi r^2 dr$.
Integrating again Eq. (\ref{kb}) with $k(0)=0$, we obtain $k(r) < 0 \ \forall r
>0$.
Accordingly, the skewness of $\Delta u_r$ and of $\partial u_r /\partial r$ are
both negative
\bea
\begin{array}{l@{\hspace{-0.cm}}l}
\ds H_3(r) = \frac{\langle (\Delta u_r)^3 \rangle}{\langle (\Delta u_r)^2
\rangle^{3/2}} 
\ds \equiv \frac{6 k(r)}{\left( 2(1-f(r))\right)^{3/2} } < 0, \ \forall r>0, 
\\\\
\ds H_3(0) = \lim_{r \rightarrow 0}H_3(r) = 
\frac{k^{III}(0)}{(-f^{II}(0))^{3/2}} <0
\end{array}
\label{kk3}
\eea
where $f = \langle u_r u_r' \rangle/u^2$ is the longitudinal correlation
function, 
and the superscript Roman numerals denote derivatives with respect to $r$. 
As $\nabla \cdot \left\langle ({\bf u}\cdot {\bf u}' ) ({\bf u}-{\bf u}')
\right\rangle$ $= O(r^2)$  near the origin \cite{Karman38}, $H_3(0)<0$ assumes
finite value, and since  
$\ds \lim_{r \rightarrow \infty} f=  \lim_{r \rightarrow \infty}k=0$, then 
$\ds \lim_{r \rightarrow \infty} H_3(r)=0$.

\bigskip

In conclusion, the proposed property of the bifurcations (\ref{cond_3}) 
implies the phenomenon of kinetic energy cascade and 
that the bifurcations are the driving force of turbulence.

\bigskip

\section{Fixed points and bifurcations of velocity fields  \label{Fixed points}}

The NS--bifurcations exhibit implications for what concerns the characteristic scales
of the velocity field. 

In the present work, to analyze this link between NS--bifurcations and characteristic scales, the fixed points associated to the current velocity field 
${\bf u}({\bf x}, t) \in C^1 \left( \{ {\bf x} \} \times \{ t \} \right)$ are
first introduced. 
\begin{figure}[h]
\centering
\vspace{-0.mm}
\hspace{-0.mm}
\includegraphics[width=0.500\textwidth]{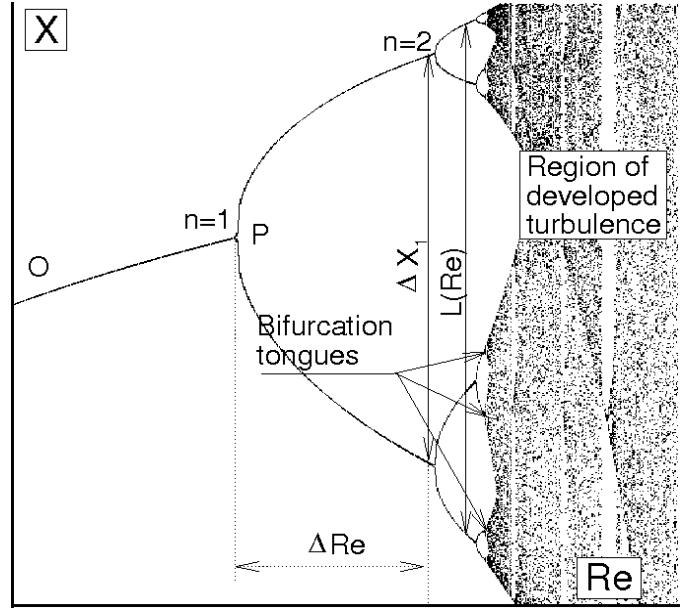}
\caption{Qualitative scheme of bifurcations of a given velocity field.}
\label{figura_2}
\end{figure}
These fixed points are here defined as the points ${\bf X}$ satisfying 
\bea
\hat{\bf u}({\bf X}; Re) = 0,
\label{bif_2}
\label{bif_1}
\eea
where $\hat{\bf u}({\bf X}; Re)$ is calculated with Eq. (\ref{inv}).
To study these points, we recall that ${\bf u} ({\bf x}, t)$ determines 
$\hat{\bf u}({\bf x}; Re) = {\bf N}^{-1} (\dot{\bf u}; Re)$ which is not unique
and depends on the Reynolds number.
Thus, these points also depend on $Re$, and if ${\bf X}(Re_0)$ 
represents the fixed points calculated at $Re_0 << Re^*$, ${\bf X}(Re)$ can be
formally obtained with the implicit function theorem
\bea
\ds {\bf X}(Re) = 
\ds {\bf X}(Re_0) +
\ds  \int_{Re_0}^{Re} {\nabla_{\bf x} \hat{\bf u}}^{-1} {\nabla_{\bf u} {\bf
N}}^{-1}  \ \frac{\partial {\bf N}}{\partial Re} \ dRe
\label{bif_map}
\eea 
where $Re > Re_0$.
${\bf X}(Re)$ can be determined with Eq. (\ref{bif_map}) if 
$\det \left( \nabla_{\bf u}  {\bf N} \ \nabla_{\bf x} \hat{\bf u} \right) \ne 0$.
If we exclude the cases where  $\det{\nabla_{\bf x} \hat{\bf u}}$= 0, the
u--bifurcations are defined as those fixed points where the operator
$\nabla_{\bf u}  {\bf N}$ admits at least one eigenvalue with zero real part.
Hence, the u-bifurcations are the image of the NS--bifurcations in the fluid
domain, and the previous considerations concerning the route toward the chaos
can be applied to Eq. (\ref{bif_map}).
For unsteady flow, the fixed points continuously vary with the time.
Figure \ref{figura_2} shows the situation corresponding to unsteady velocity
fields at a given instant $\bar{t}$, where $\Delta X$ is the bifurcation scale, a
characteristic length which expresses the distance between branches which born
from the same bifurcation.

These u--bifurcations determine sizable variations on the characteristic scales
of the velocity field. If, for $Re<<Re^*$, the velocity field is represented by 
the Fourier series of a given basic scale, one bifurcation corresponds 
to new solutions $\hat{\bf u}$ whose Fourier characteristic lengths,
about proportional to $\Delta X$, are independent from the previous one. 
Therefore, each bifurcation adds new independent scales, 
and, after the transition, the several characteristic lengths are continuously 
distributed, thus $\bf u$ is there represented by the Fourier transform.

Returning to Eq. (\ref{bif_map}), the diagram of Fig.\ref{figura_2} is qualitatively similar to that of Fig. \ref{figura_1}, therefore it exhibits the same general properties:
1) $L(Re)$ smoothly varies with respect to $Re$, where $L(Re)$ is the overall
dimension  of the bifurcation tree along $X$. In particular, we assume that $L(Re)$ is
a rising function of $Re$ with $L(Re)$ $\in$ $C^0(Re)$. 
2) existence of bifurcations tongues whose width increases with $Re$, and that
thereafter are overlapped.

Another important property is that, for $Re < Re^*$, the bifurcations are
limited in number, and the sum of the distances between contiguous branches does
not exceed $L(Re)$ \cite{Mandelbrot02, Mainzer05}
\bea
\ds \Sum_{n \ne 1 } \Delta X_n < \Delta X_1 - \Delta X_N < L(Re)
\label{s1}
\eea

 Vice versa, for $Re > Re^*$, the bifurcations frequentely happen, the
bifurcations tree will exhibit fractional dimension and self--similarity 
\cite{Mandelbrot02, Mainzer05}, while the distance between the successive
 u--bifurcations is very small. Accordingly, we have 
\bea
\ds \Sum_{n \ne 1 } \Delta X_n > L(Re) > \Delta X_1
\label{s2}
\eea
This equation expresses the well known property that the perimeter $\cal P$ of
one fractal geometry is much greater than its overall dimension as 
${\cal P} \approx A^{D/2}$ and $L(Re) \approx A^{1/2}$, where $A$ 
is the area of the fractal object and $D$ is its fractal dimension
(see for instance Ref.  \cite{Mandelbrot02}  and references therein).

Hence, immediately before the transition ($Re\lesssim Re^*$), 
the situation is characterized by a distribution of discrete scales 
$\Delta X_n$,  whereas for $Re > Re^*$, the bifurcations behave like continuous transitions
and $\Delta X$ plays  the role of a real variable.

\bigskip

\section{Bifurcations cascade in terms of length scales \label{length scales}}

\begin{figure}[h]
\centering
\vspace{-0.mm}
\hspace{-6.0mm}
\includegraphics[width=0.4 \textwidth]{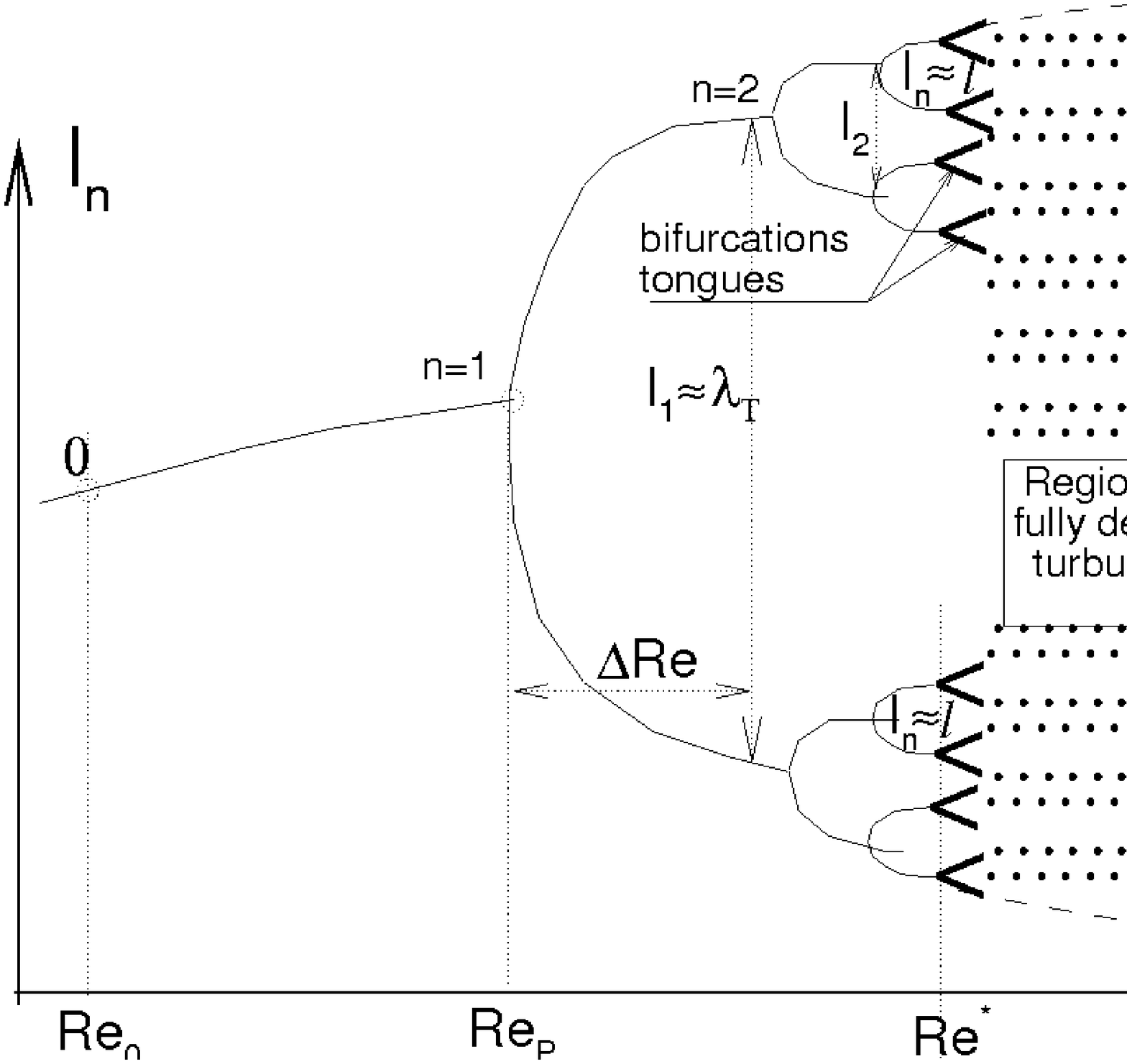}
\qquad\qquad
\includegraphics[width=0.4 \textwidth]{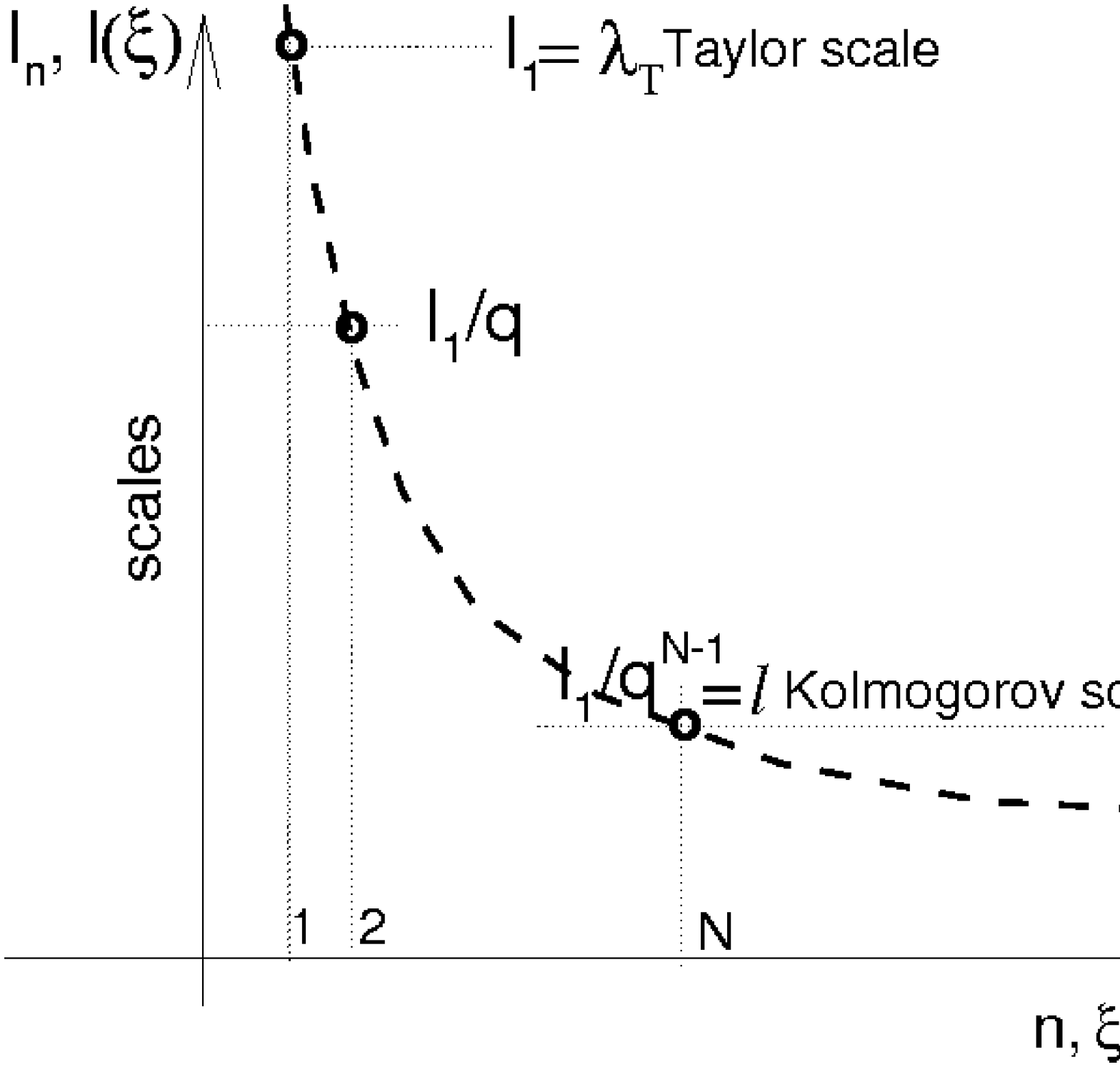}
\caption{Left: Schematic of the bifurcations cascade law in terms of length
scales.
Right: Discrete (Filled symbols) and continuous (dashed line) length scales
distribution 
for $Re\approx Re^*$ }
\label{figura_3}
\end{figure}

The length scales vary with time and their characteristic values 
are expressed in function of $n$.
Figure \ref{figura_3} (Right) qualitatively shows $l_n$ immediately before the
transition ($Re \lesssim Re^*$, filled symbols), where $N$ is the number
of encountered bifurcations at $Re^*$.
These scales, discretely distributed, are expressed by the succession 
\bea
\left\lbrace {\mbox l}_n , \ n=1, 2,...N \right\rbrace
\eea

Conversely, to represent the continuous scales $l$ in developed turbulence, $l$
is in terms of $\xi>$0, a real variable which replaces $n$ in developed turbulence.
Hence
\bea 
\ds l = l(\xi), \ \ \xi \in \mathbb{R}^+
\eea
where $l(\xi)$ $d \xi$ represents the elemental scale in developed turbulence.

As discussed in sect. \ref{Fixed points}, for $Re \gtrsim Re^*$,
the lengths  describe a dense set which includes the characteristic
scales before the transition.
Accordingly, the function $l(\xi)$ is chosen in such a way that
\bea
\ds l(n)={\mbox l}_n, \ n=1, 2, ..., N
\label{pre_l}
\eea
Next, because of the aforementioned self--similarity 
\cite{Mandelbrot67, Mandelbrot02}, ${\mbox l}_n$ and $l(\xi)$  
are supposed to be, respectively, a geometric progression and an exponential function, 
i.e.
\bea 
\begin{array}{l@{\hspace{-0.cm}}l}
\ds {\mbox l}_n = \frac{{\mbox l}_1} {q^{n-1}}, \ \ n=1, 2,..., N, \ \mbox{for}
\ Re \lesssim Re^*  \\\\
\ds l(\xi) = \frac{{\mbox l}_1} {q^{\xi-1}}, \ \ \xi \in \mathbb{R}^+, \
\mbox{for} \ Re \gtrsim  Re^* 
\end{array}
\label{scales0}
\eea
where $q > 1$.

Now, in order to obtain the estimation of $q$, observe that following the
inequality (\ref{s1}), the sum of the distances $\Delta X_n$ does not exceed 
$\Delta X_1 - \Delta X_N$, for $Re \lesssim Re^*$, i.e. 
\bea
\begin{array}{l@{\hspace{-0.cm}}l}
\ds {\Sum_{n=2}^{N}} {{\mbox l}_n}
\ds  <  {{\mbox l}_1} -{{\mbox l}_N} \equiv  {{\mbox l}_1} \left(
1-\frac{1}{q^{N-1}}\right)  
\end{array}
\label{int0}
\eea
On the contrary, for $Re > Re^*$, thanks to the fractal properties of the
bifurcations, the sum of such these distances is much greater than $L(Re)$
\cite{Mandelbrot02}(see ineq. (\ref{s2})), and this can be expressed taking
into account that $\xi \in \mathbb{R}^+$
\bea
\ds \int_1^\infty l(\xi) d \xi 
> {\mbox l}_1 
\label{int1}
\eea
As $\{ {\mbox l}_n, n =1, 2,... \}$ is a geometric succession, 
the inequality (\ref{int0}) is satisfied for $q > 2$ and $N$ arbitrary, whereas
the condition (\ref{int1}) is satisfied for $q < e$, that is
\bea
2 < q < e
\eea

\bigskip

\section{Estimation of the critical Taylor--scale Reynolds number
\label{Estimation re}}

In fully developed turbulence, the Taylor--scale Reynolds number is defined by
\bea
\ds R_\lambda  =  \frac{u \lambda_T}{\nu}
\eea 
where  $\lambda_T = {1}/{\sqrt{-f^{II}(0)}}$ is the Taylor scale.
$R_\lambda$, $\lambda_T$ and $u$ are linked
by means of the relation \cite{Batchelor53}
\bea
\ds \frac{\lambda_T}{\ell} = 15^{1/4} \sqrt{R_\lambda}, \ \ \ \ 
\label{scales}
\eea
where $\ell$ is the Kolmogorov microscale.

The critical Taylor--scale Reynolds number $R_{\lambda}^*$ is first estimated 
starting from the route toward the chaos, 
using the bifurcations cascade seen at the previous section, and assuming
opportune properties of the length scales. 
Thereafter, $R_{\lambda}^*$ is also estimated beginning from the fully developed
isotropic  turbulence, adopting the closure equation presented in
Ref. \cite{deDivitiis_1} for the von K\'arm\'an--Howarth equation, and a plausible
condition for $\lambda_T$.

\bigskip

\subsection{Estimation of $R_\lambda^*$ through the route toward the chaos}

To estimate $R_\lambda^*$ through the route toward the turbulence, the
relationship 
between $R_\lambda^*$ and $N$ is searched.
Now, to obtain this link, observe that, due to the presence of the bifurcation 
tongues, the Kolmogorov scale $\ell$ can not exceed $l(N)$ at the onset of
turbulence.
At the same time, $L(Re) \approx \lambda_T$ for $Re>Re^*$, 
and thanks to the smooth variations of $L(Re)$ through the transition 
(see Fig. \ref{figura_3} (Left)),  $\lambda_T$ can not be less than
$l(1)$, thus $\ell < l(n) < \lambda_T$ for $Re \approx Re^*$. 
Hence, according to Eq. (\ref{scales0}), we assume that 
\bea
\ds l(n) = \frac{\lambda_T}  {q^{n-1}}, \ \ \  \ell = \frac{\lambda_T} 
{q^{N-1}}
\label{scales01}
\eea
Combining Eqs. (\ref{scales01}) and (\ref{scales}), we obtain 
\bea
R_\lambda^* = \frac{q^{2 N -2}}{\sqrt{15}}
\label{scales1}
\eea
which expresses the searched relationship.
With reference to table \ref{table1}, all the values of $R_\lambda^*$ calculated
for $N=2$ and $q \in [2, e]$ are of the order of the unity and this is not
compatible with $\lambda_T$ which represents the correlation scale, while the
values $R_\lambda^*= 4 \div 14$ obtained for $N=$ 3 and $q \in [2, e]$ are
acceptable. 
In particular, if $q$ is assumed to be equal to the second Feigenbaum constant 
($\alpha = 2.502...$), $R_\lambda^* \simeq$ 10.
For $N=4$, all the values of $R_\lambda^*$ seem to be quite high in comparison
with a plausible 
minimum values of $R_\lambda$, expecially for high values of $q$.
\begin{table}[t]
\centering
\caption{Critical Taylor--scale Reynolds number 
calculated for $N=$ 2, 3, and 4, for different values of $q$. }
\vspace{2. mm}
\begin{tabular}{cccc} 
\hline
\hline
$q$   &  $Re_\lambda^*$(N=2)&  $Re_\lambda^*$(N=3)  & $Re_\lambda^*$(N=4)  \\
\hline 
2.000                   & 1.03 &    4.13 & 16.52    \\
2.250                   & 1.31 &    6.62 & 33.50    \\
$\alpha$ $\simeq$ 2.503 & 1.62 &    10.13 & 63.49   \\
e $\simeq$ 2.718        & 1.91 &    14.10 & 104.16   \\ 
\hline
\hline
 \end{tabular}
\label{table1}
\end{table} 

These orders of magnitude of $R_\lambda^*$ calculated for $N=3$, agree with the
different theoretical routes to the turbulence \cite{Ruelle71, Feigenbaum78,
Pomeau80, Eckmann81}, and with the diverse experimental data \cite{Gollub75,
Giglio81, Maurer79} which state that the transition occurs when $N \gtrsim 3$. 

\bigskip

\subsection{Estimation of $R_\lambda^*$ through the fully developed turbulence}

Next, to estimate $R_\lambda^*$ starting from the regime of fully developed
homogeneous isotropic turbulence, the solutions of the von K\'arm\'an--Howarth
equation are analyzed in function of 
$R_\lambda$. To determine $R_\lambda^*$, we need an auxiliary condition which
defines the lower limit for the existence of this regime of turbulence.
To found this condition, observe that the homogeneous isotropic turbulence is an
unsteady 
regime, where $u$ and $\lambda_T$ change with $t$ according to Refs. \cite{Karman38,
Batchelor53}
\bea
\begin{array}{l@{\hspace{-0.cm}}l}
\ds \frac{d u^2}{d t} = - \frac{10 u^2 \nu}{\lambda_T^2}
\end{array}
\label{r0}
\eea
\bea
\begin{array}{l@{\hspace{-0.cm}}l}
\ds  \frac{5  \nu}{\lambda_T^4}  + \frac{1}{\lambda_T^3} 
\frac{d \lambda_T}{d t}   = 
\frac{7}{6} u k^{III}(0) + \frac{7}{3} \nu  f^{IV}(0)
\end{array}
\label{r2}
\eea
where Eqs. (\ref{r0}) and (\ref{r2}) are the equations for the coefficients of
the powers
$r^0$ and $r^2$, respectively, of the von K\'arm\'an--Howarth equation.
The term responsible for the energy cascade is the first one at the R.H.S. of
Eq. (\ref{r2}),
whereas the second one is due to the viscosity.
According to Eq. (\ref{r2}), if the energy cascade is sufficiently stronger than
the viscosity effects, then ${d \lambda_T}/{d t} <0$.
Hence, a reasonable condition to estimate $R_\lambda^*$ can consist in to
search the value of $R_\lambda$ for which 
\bea
\ds \frac{d \lambda_T}{dt} =0 
\label{lt=0}
\eea
at a given instant.
This value of $R_\lambda^*$ depends on the adopted closure equation for $K$.
If we use the results of the Lyapunov theory proposed by Ref. \cite{deDivitiis_1},
$K$ is in terms of $f$ and ${\partial f}/{\partial r}$
\bea
\ds K = u^3 \sqrt{\frac{1-f}{2}} \frac{\partial f}{\partial r}
\eea
thus Eq. (\ref{lt=0}) is satisfied for \cite{Batchelor53}
\bea
\ds R_\lambda \equiv R_\lambda^* = 2 \left( \frac{7}{3} \varphi -5 \right)  
\ \mbox{where} \ \varphi = \frac{f^{IV}(0)}{\left( f^{II}(0)\right)^2 }
\eea 
Following such estimation,  $R_\lambda^*$ is related to the behavior 
of $f$ near the origin through $\varphi > 15/7$.
For instance, when $f$ is a gaussian function 
\bea
\ds f= \exp \left( f^{II}(0) \frac{r^2}{2} \right), \ \mbox{then}  \ \varphi =3,
\ \ R_\lambda^* =4.
\eea
whereas if, according to the Kolmogorov law, $f$ behaves like
\bea
\ds f  \approx 1-  c r^{2/3}, \ \ \ c >0, \ \mbox{then}  \ \varphi =4.8, \ \
R_\lambda^* =12.4.
\eea
where ${f^{I}(\lambda_T/\sqrt{2})}$ and ${f^{II}(\lambda_T/\sqrt{2})}$ are
assumed to be equal to the corresponding derivatives of $1+1/2 f^{II}_0 r^2+1/4!
f^{IV}_0 r^4$ in $r=\lambda_T/\sqrt{2}$.
These values are in qualitatively good agreement with those of the previous
analysis based on the bifurcations.

\bigskip

\section{\bf Conclusion \label{Conclusion}}

We conclude this work by observing that the proposed properties of 
fully developed turbulence based on bifurcations
explain the energy cascade phenomenon in agreement 
with the literature, and motivate the fact that the local fluid strain can be
much faster than the fluid state variables. 
Furthermore, the spatial representation of the bifurcations justifies that 
the bifurcations cascade can be expressed in terms of length scales, and 
allows to argue that the scales are continuously distributed in developed 
turbulence.
The fractal properties of the bifurcations and the proposed link between 
the characteristic scales through the transition lead to a relationship
between critical Reynolds number and number of bifurcations at the transition,
resulting $N$=3 and $R_\lambda^* \approx 4 \div 14$ in line with the literature.
$R_\lambda^*$ is also estimated as that value of the Taylor--scale Reynolds
number at which $d \lambda_T/dt =0$ in the isotropic turbulence. 
The two procedures provide values of $R_\lambda^*$ in agreement with each other.

\bigskip

\section{\bf Acknowledgments}

This work was partially supported by the Italian Ministry for the Universities 
and Scientific and Technological Research (MIUR).

\bigskip

\end{document}